\documentclass{WileyMSP-template}
\usepackage{gensymb}
\usepackage{tabularx}
\usepackage{graphicx}
\newcommand\sbullet[1][.5]{\mathbin{\vcenter{\hbox{\scalebox{#1}{$\bullet$}}}}}
\newcommand\bb{$\sbullet$~}

\begin{document}

\pagestyle{fancy}
\rhead{\includegraphics[width=2.5cm]{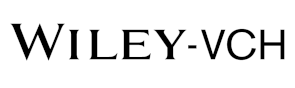}}

\title{Hybrid quantum systems with artificial atoms in solid state}

\maketitle

\author{Cleaven Chia$^1$},
\author{Ding Huang$^1$},
\author{Victor Leong$^1$},
\author{Jian Feng Kong$^2$},
\author{Kuan Eng Johnson Goh$^{1, 3, 4*}$}

\begin{affiliations}
${}^1$ {\footnotesize A*STAR Quantum Innovation Centre (Q.InC), Institute of Materials Research and Engineering (IMRE), Agency for Science, Technology and Research (A*STAR), Singapore, 138634, Republic of Singapore}

${}^2$ {\footnotesize A*STAR Quantum Innovation Centre (Q.InC), Institute of High Performance Computing (IHPC), Agency for Science, Technology and Research (A*STAR), Singapore, 138632, Republic of Singapore}

${}^3$ {\footnotesize Division of Physics and Applied Physics, School of Physical and Mathematical Sciences, Nanyang Technological University, Singapore, 639798, Singapore}

${}^4$ {\footnotesize Department of Physics, National University of Singapore; 2 Science Drive 3, Singapore, 117551, Singapore}

\end{affiliations}
\vspace{5mm} 

{\footnotesize *Corresponding Author Email: kejgoh@yahoo.com}



\begin{abstract}
\justifying
The development of single-platform qubits, predominant for most of the last few decades, has driven the progress of quantum information technologies but also highlighted the limitations of various platforms. Some inherent issues such as charge/spin noise in materials hinder certain platforms, while increased decoherence upon attempts to scale-up severely impact qubit quality and coupling on others. In addition, a universal solution for coherent information transfer between quantum systems remains lacking. By combining one or more qubit platforms, one could potentially create new hybrid platforms that might alleviate significant issues that current single platform qubits suffer from, and in some cases, even facilitate the conversion of static to flying qubits on the same hybrid platform. While nascent, this is an area of rising importance that could shed new light on robust and scalable qubit development and provide new impetus for research directions. Here, we define the requirements for hybrid systems with artificial atoms in solid state, exemplify them with systems that have been proposed or attempted, and conclude with our outlook for such hybrid quantum systems.

\end{abstract}


\justifying
\section{Introduction}

Quantum technologies are motivated by the potential of quantum systems to outperform their classical counterparts: qubits are able to store more information compared to classical bits by virtue of superposition and entanglement allowing potential computational speed-ups in solving certain problems \cite{nielsen2011quantum}, and quantum systems can act as sensors due to their intrinsic sensitivity to the environment. The development of quantum processors and sensor technologies have focused on single-platform qubits for a large part of the last three decades. This is due to the significant challenge of realizing even a single qubit on any feasible platform \cite{diVincenzo2000thephysical}, and also to the consideration of practical scalability of such qubits on a single platform. Many single-platform qubit flavors are now in existence at various levels of maturity, e.g. trapped ions, neutral atoms, electron/nuclear spins, superconducting transmon qubits, color centers, photons, and even nanomechanical oscillators. A number of these systems have had fast developmental ramp-ups owing to mature technological materials/components/techniques being available, e.g. GaAs spin-qubits, superconducting qubits, ion-trap qubits, photonic qubits, but most of these platforms have found or are beginning to find plateaus in their developments. Some are so because of intrinsic issues such as charge or spin noise related to the platform, others because of the unwanted external sources of decoherence that increases significantly upon attempts to scale up to multiple qubits\cite{deLeon2021materials}.

\vspace{5mm} 

To overcome these challenges, hybrid systems that couple one or more qubit platforms have been devised. The motivation behind developing hybrid quantum systems lies in using the strengths of one qubit platform to overcome the issues of another qubit platform, thereby outperforming single-platform qubits. One example where the need for hybrid systems is in the field of quantum information processing, where the simplest architecture can be thought of as consisting of nodes that store quantum information, and channels that connect and transmit quantum information between nodes \cite{kimble2008quantum}. These differing criteria are satisfied by different modalities. Nodes typically comprise of stationary qubits, such as spins or superconducting qubits, that are able to maintain coherence over long times. However, scaling up to many nodes can result in increased crosstalk in the case of superconducting qubits \cite{deLeon2021materials}, or decoherence induced by neighboring spins \cite{wolfowicz2021quantum}. As such, modular architectures have been proposed, where small modules containing few nodes are connected to other spatially-separated modules via flying qubits \cite{awschalom2021development}. Such connections can be realized by using propagating photons as a channel. In particular, optical photons at telecommunications wavelengths are desirable as a channel due to their low attenuation in optical fibers and their compatibility with existing fiber optic infrastructure, enabling long-range interactions that allow for large-scale quantum processors and distributed quantum networks to be realized \cite{awschalom2021development}.

\vspace{5mm} 

Another example requiring hybrid quantum systems is in the detection of single spins: due to the weak magnetic dipole moment of spins, direct readout of spin states using magnetic fields is challenging \cite{rugar2004single}. As such, spin states are often readout through more sensitive mechanisms in hybridized systems: e.g. through spin-to-charge conversion followed by electrical readout of photocurrent \cite{wolfowicz2021quantum}; or through spin-photon interfaces wherein optical transitions can be used to resolve spin states, with the emitted photons detected through photodetectors such as avalanche photodiodes (APDs) or superconducting nanowire single-photon detectors (SNSPDs) \cite{heinrich2021quantum, kim2020hybrid}. As most long range quantum information is currently transmitted via photons, the conversion from spin to photon states is often regarded as an enabler for static to flying qubit realization in practical systems.

\vspace{5mm} 

A third example requiring hybrid quantum systems lies in converting quantum states to a modality that preserves such states for a longer time, such as use of nuclear spins as an intermediary to store quantum information encoded in neighboring electron spins in diamond \cite{pompili2021realization} or in silicon \cite{kane1998silicon}. This is effective due to the insensitivity of nuclear spins to environmental noise giving them long coherence times. Additionally, the use of quantum memories have been shown to improve the sensitivity of a quantum sensor, by prolonging the duration over which signal is accumulated \cite{zaiser2016enhancing}.

\vspace{5mm} 

The above applications all require the use of a localized system in which information can be encoded, or with which a physical signal can be sensed. One such system is artificial atoms -- localized systems with discrete energy levels that can be engineered through design or via their local environment \cite{buluta2011natural}. This is in contrast to natural atoms (both neutral and ions), whose energy levels are fixed and identical by virtue of their atomic species. The size of artificial atoms can span from the atomic scale that includes dopants (e.g. phosphorus dopants in silicon) and vacancy complexes (nitrogen and silicon vacancies in diamond), to the macroscopic scale of superconducting qubits. The solid-state nature of artificial atoms precludes the need for levitation or trapping, and are compatible with commercial fabrication methods. This, combined with their designer energy levels, make them an attractive component for hybrid quantum systems. 

\vspace{5mm} 

In this article, we will review hybrid solid-state systems that have been realized for the fields of quantum computation, communications and sensing,
where their compact footprint and ease of integration can potentially be beneficial for production scale-up leading to large-scale deployment for relevant applications. We develop a set of criteria for evaluating different hybrid solid-state systems and provide guidance for areas which can benefit from such hybrid quantum systems.

\section{Evaluating hybrid quantum systems}
How do we evaluate hybrid quantum systems? While metrics such as quality factors, bandwidth, spatial and spectral resolution are commonly used, these are focused on specific device types and applications. In this section, we first briefly review the major applications of quantum systems in the areas of quantum computing, communications, and sensing. We then present a set of requirements that provides a common rubric to assess hybrid quantum systems, and how the hybridization or coupling of different systems enhances their performance and suitability in developing practical quantum devices for these applications.

\subsection{Quantum Computation} \label{Section:QC}
Quantum computation utilizes entanglement and superposition in quantum mechanical systems to perform computations. Generally, quantum information is encoded within qubits, and operations are performed on the qubits to manipulate their states and obtain the results of a calculation. As qubits can be prepared in a superposition of its two basis states, this enables more complex computations compared to classical computers, in which information can only be encoded in one of two discrete states of classical bits. In this work, we shall focus on discrete variable quantum computation which utilizes qubits as computing elements.

\vspace{5mm} 

\begin{figure}[thb]
\centering
\includegraphics[width=0.9\textwidth]{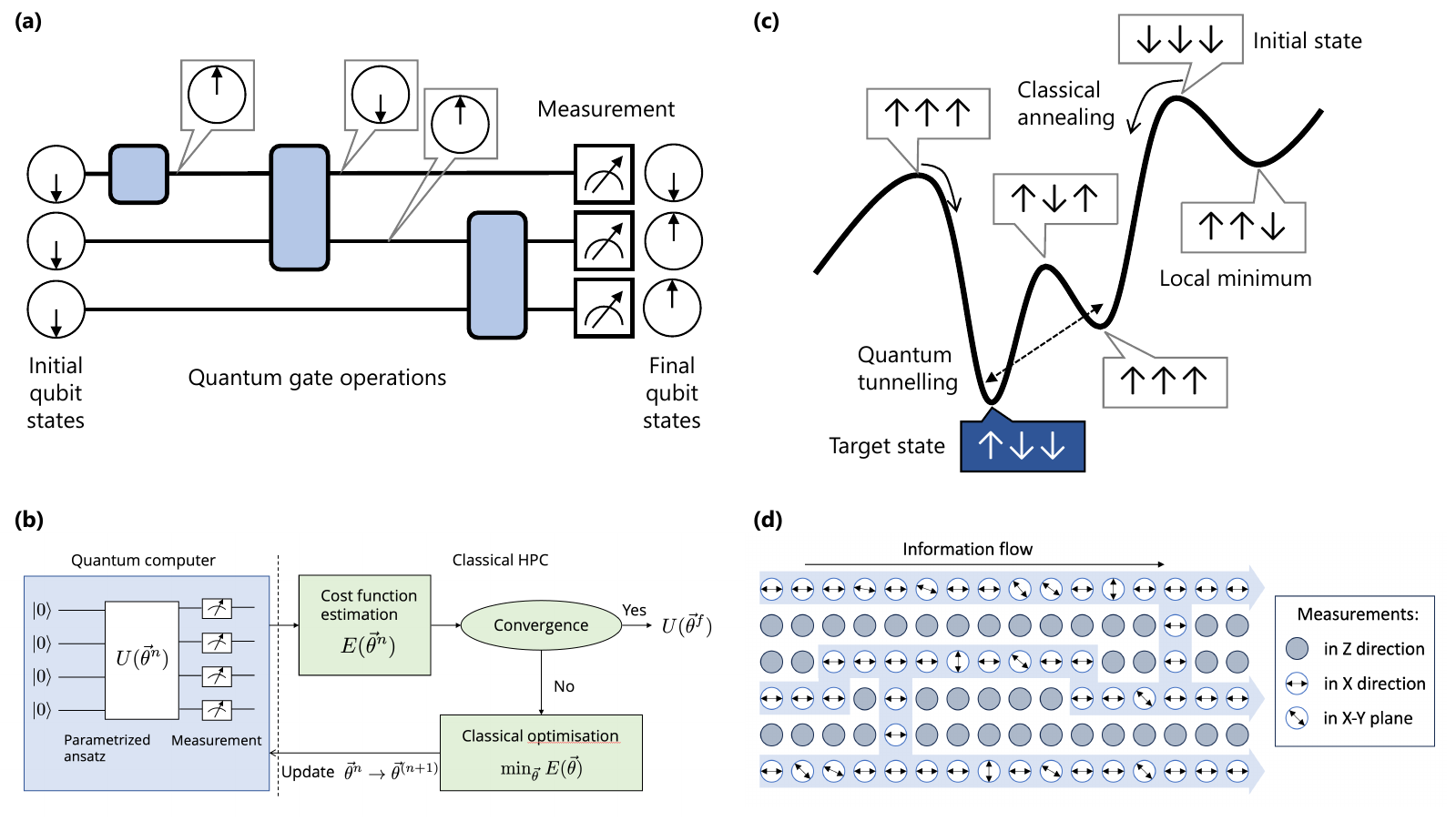}
\caption{Models of quantum computing. (a) Gate-based quantum computing with gate operations on one or more qubits. (b) Variational quantum eigensolver (VQE) for current noisy intermediate-scale quantum (NISQ) devices. (c) Quantum annealing, with different states at different points in the energy landscape. (d) Measurement-based quantum computing (MBQC) illustrated with a sequence of single-qubit measurements performed on the initial resource state.}
\label{fig:qcomputingmodels}
\end{figure}

Within quantum computation, there are various models amongst which the most well-known is the gate-based model. In close analogy to its classical counterpart, computations are performed by applying a series of logic gates on an initialized qubit state as depicted in \textbf{Figure \ref{fig:qcomputingmodels}}(a). It has been shown that any set of gate operations can be expressed in terms of a universal set of single- and two-qubit gates \cite{nielsen2011quantum}. Current hardware implementations with limited qubit numbers and gate fidelities still do not allow for universal fault-tolerant quantum computing. The Variational Quantum Eigensolver (VQE) is a hybrid algorithm that combines quantum and classical computing techniques to determine the lowest eigenvalue and the corresponding eigenvector of a Hamiltonian -- a crucial task in fields like quantum chemistry~\cite{peruzzo2014variational,mcclean2016theory}. This process involves a quantum circuit with adjustable parameters, which are optimized through a classical optimization loop shown in Figure~\ref{fig:qcomputingmodels}(b). In this loop, the expectation value of the Hamiltonian, derived from the quantum circuit's measurement results, is iteratively minimized. The effectiveness of VQE heavily relies on the choice of the quantum circuit, and determining the optimal circuit for a given problem typically requires extensive benchmarking. By using low-depth circuits, VQE can be well implemented on current Noisy Intermediate-Scale Quantum (NISQ) devices. As quantum hardware improves, more complex circuits can be implemented, potentially allowing VQE to tackle more complex and larger-scale problems~\cite{bharti2022noisy}.

\vspace{5mm} 

Another model of quantum computation is quantum annealing, in which a set of qubits is evolved with time in analogy to classical annealing, so as to find their lowest-energy configuration as a solution to the computational problem \cite{das2008colloquium, albash2018adiabatic}. This is shown schematically in Figure \ref{fig:qcomputingmodels}(c).

\vspace{5mm} 

A third model is measurement-based quantum computation, which starts with a highly-entangled multi-qubit state on which successive measurements are performed in order to reach a computation result \cite{raussendorf2001one, raussendorf2002computational, raussendorf2003measurement, briegel2009measurement}. This model requires a large number of entangled qubits to form the initial resource state and is sometimes called the ``one-way" method as the highly entangled input is destroyed by the measurement. As the computational outcome is random due to the quantum nature, the procedure involves a series of measurement of ancillary qubits and output correction before converging on the final result (Figure~\ref{fig:qcomputingmodels}(d)).

\vspace{5mm} 

Regardless of the model of quantum computation employed, there are some common essential components across the different models. Such components include quantum memories to store quantum information for sufficiently long times between computations; quantum processors that manipulate qubit states through entanglement and/or measurement to perform computations; and quantum interconnects between memories and processors that enable larger-scale computations \cite{perez2011quantum, awschalom2021development}. The requirements for quantum memories, processors and interconnects differ: while memories require long storage and coherence times (relative to computation times), processors and interconnects need to perform gate and entangling operations quickly so as to minimize total computation times. In addition, processors and interconnects should transfer qubit states coherently between modalities so as to preserve quantum information; in particular, quantum state transfer needs to remain coherent over longer distances for quantum interconnects compared to quantum processors. Due to the disparate requirements of quantum memories, processors and interconnects, and no single material platform has yet been shown to satisfy these requirements simultaneously, hence hybrid quantum systems are sought for integrating these demands~\cite{elshaari2020hybrid}.

\subsection{Quantum Communications} \label{Section:QComm}

\begin{figure}[thb]
\centering
\includegraphics[width=0.85\textwidth]{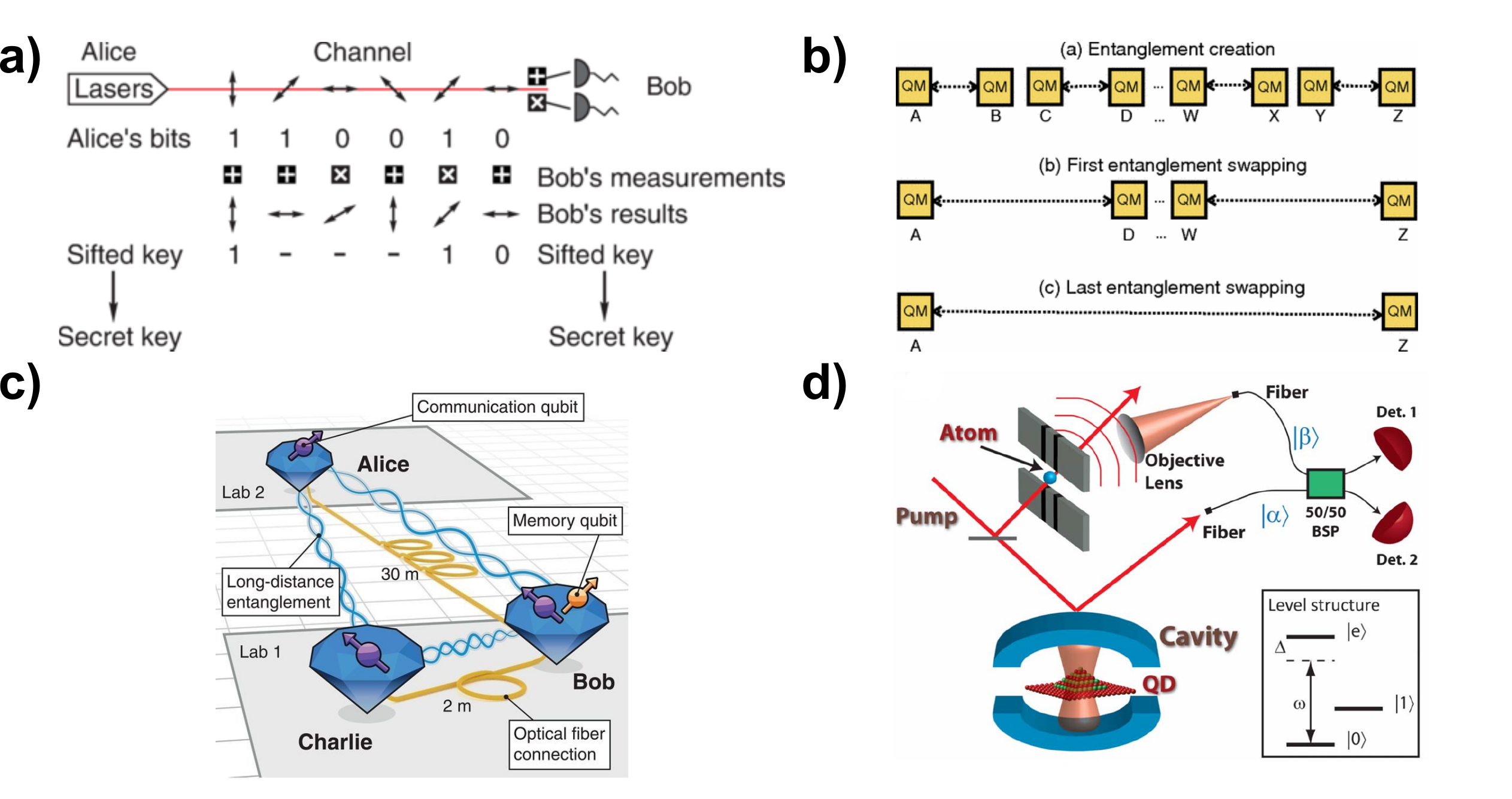}
\caption{Schematic illustration of Quantum Communications protocols. (a) Direct transmission of quantum information using the Bennett-Brassard 1984 protocol. The information is encoded in the polarization states of single photons and used as the secret keys for communications. Reprinted with permission\cite{xu2020secure}. Copyright 2020, American Physical Society. (b) Quantum repeater protocol. To create an entanglement pair distributed over long distances (e.g. in between Node A and Node Z), entanglement is first created independently within short elementary links between Node A and B, Node C and D etc. Entanglement links can then be swapped in between neighboring nodes and longer links can be created in between Node A and D, Node W and Z etc. Concatenated entanglement swapping will eventually create an entanglment link between Node A and Z. Reprinted with permission\cite{sangouard2011quantum}. Copyright 2011, American Physical Society. (c) Three-node quantum network based on NV centers in diamond. Reprinted with permission\cite{pompili2021realization}. Copyright 2021, The American Association for the Advancement of Science. (d) Proposed scheme to realize entanglement between a single atom and an optically-active quantum dot. This could be used to build hybrid quantum network nodes with different memory qubits. Reprinted with permission\cite{waks2009protocol}. Copyright 2009, American Physical Society.}
\label{fig:QComm}
\end{figure}

Quantum communications promises secure exchange of information that is guaranteed by the fundamental laws of physics rather than computational complexity \cite{bennett2014quantum, shor2000simple, gisin2002quantum} (\textbf{Figure \ref{fig:QComm}}(a)). Due to the losses associated with directly transmitting a quantum state, most of the protocols for quantum communications requires the generation and distribution of entangled states between distant quantum nodes \cite{kimble2008quantum,wehner2018quantum,xu2020secure}. 
\vspace{5mm} 

Similar to classical communications, quantum communications schemes encode the information in optical photons and use the photons to transmit quantum information over a long distance. However, the photon loss in the transmission channels increases exponentially with distance. In classical communications, this can be overcome by straightforward signal amplifications. Due to the no-cloning theorem, this could not be achieved in quantum communications. A possible solution to overcome the photon loss is the use of intermediate nodes, known as quantum repeaters \cite{briegel1998quantum, duan2001long}. At these intermediate nodes, the information carried by photons can be stored by the spin-wave excitation in ensembles of atoms \cite{yuan2008experimental, jing2019entanglement, yu2020entanglement}, the nuclear or electronic spin degree of freedom in single atoms \cite{ritter2012elementary, hofmann2012heralded, rosenfeld2017event}, trapped ions \cite{moehring2007entanglement, olmschenk2009quantum, maunz2009heralded, hucul2015modular,stephenson2020high}, solid-state defects \cite{bernien2013heralded, christle2015isolated, rose2018observation, pompili2021realization, anderson2022five, hermans2022qubit, stas2022robust, knaut2023entanglement}, rare-earth ions in solid-state hosts \cite{zhong2017nanophotonic,dibos2018atomic,chen2020parallel,lago2021telecom,liu2021heralded,ruskuc2022nuclear,stevenson2022erbium} and quantum dots \cite{delteil2016generation,stockill2017phase}. Entangled states over a longer distance can be created from shorter, elementary entanglement links between the intermediate quantum repeater nodes \cite{sangouard2011quantum}. As a result, the quantum repeater nodes can extend the maximum distance over which entangled states can be distributed (Figure \ref{fig:QComm}(b)). 

\vspace{5mm} 

As any implementation of a quantum repeater requires the coherent interaction between single photons and the spin degree of freedom of a stationary memory qubit, it would already satisfy our definition of a hybrid quantum system. Here, we will instead focus on hybrid implementations of quantum repeaters which consist of more than one physical system. 

\vspace{5mm} 

Despite the recent progress on both theoretical and experimental fronts, a scalable quantum repeater network that can span across the globe has not yet been demonstrated. The current implementations of quantum repeater nodes suffer from low photon-memory entanglement efficiency \cite{wei2022towards}, limited memory storage times \cite{wehner2018quantum, labay2023reducing}, low repetition rate \cite{van2020extending}, degraded performance when incorporating into nanophotonic devices \cite{ruf2019optically}, or the requirements to perform subsequent frequency conversions of the photons in order to attain low-loss transmission with optical fibers \cite{dreau2018quantum, weber2019two, singh2019quantum, schafer2023two}. 

\vspace{5mm} 

Among various implementations of quantum repeater nodes, color centers in diamond has established itself as a promising candidate due to the combination of optical addressability with long spin coherence times. Recent proof-of-principle demonstrations of small-scale quantum networks using color centers in diamond (Figure \ref{fig:QComm}(c)) include memory-enhanced quantum communication \cite{bhaskar2020experimental}, qubit teleportation between non-neighbouring nodes \cite{pompili2021realization}, extended memory storage times \cite{bartling2022entanglement, stas2022robust}, and long-distance entanglement with metropolitan fiber networks \cite{hensen2015loophole, ruf2021quantum, knaut2023entanglement}. While such results are promising, the necessity of etching bulk diamond is both practically challenging and potentially damaging to the color center's properties due to fabrication-induced damages or surface states. In Section \ref{Section:hybrid_cavity}, we discuss a potential way to circumvent this challenge using hybrid quantum systems.

\vspace{5mm} 

On a broader scale, quantum repeater nodes that consist of different memory qubits could also be considered as hybrid quantum systems. As each physical systems have their unique advantages and limitations for realizing quantum repeater nodes, a hybrid quantum repeater node that consists of entangled quantum memories in different physical systems could potentially reap the benefits of constituent memory qubits in practical quantum communications. Pioneering experiments along this direction have been done with single atoms and Bose-Einstein condensates \cite{lettner2011remote}. More recently, experiments with solid-state systems have shown the control of the internal states of a trapped ion using single photons generated by a quantum dot \cite{meyer2015direct}, and quantum-state transfers from an ensemble of atoms to rare-earth ions in solid-state hosts \cite{maring2017photonic}. Although there have been a few theoretical proposals for entanglement generation between a single trapped ion and spins in solid-state systems \cite{waks2009protocol, lilieholm2020photon} (Figure \ref{fig:QComm}(d)), such hybrid entanglement has not been demonstrated yet. More work is needed in this area to show the advantages of a hybrid quantum network with different physical systems.  

\vspace{5mm} 

In addition, we also note that entangled states shared between disparate physical systems could be used to implement teleportation-based quantum gate operations for modular quantum computation \cite{awschalom2021development}. Hybrid entanglement with solid-state spin qubits could potentially facilitate the connections between small-scale neutral-atom quantum processors based on Rydberg interactions or trapped-ion quantum processors. 

\subsection{Quantum Sensing}

Quantum sensing can be broadly defined as either i) using a quantum object or system to measure a physical quantity, 
or ii) harnessing quantum mechanical properties, e.g. entanglement, to enhance sensing performance beyond classical limits~\cite{degen2017quantum}.
Here, the premise of this work largely limits our discussion to the former case, specifically the use of solid-state artificial atoms as quantum sensors,
which is actively being developed in a number of systems. 
Nitrogen vacancy (NV) centers in diamond are prominent for their applications especially in magnetometry~\cite{balasubramanian2008nanoscale,wang2022picotesla},
but also in sensing electric fields~\cite{michl2019robust,bian2021nanoscale}, temperature~\cite{neumann2013high,choe2018precise} and strain~\cite{doherty2014electronic,ho2023spectroscopic}.
Other examples of such sensors include dopants in silicon~\cite{pla2018strain}, atomic-scale defects in silicon carbide (SiC)~\cite{jiang2023quantum,castelletto2023quantum}, and superconducting qubits~\cite{bal2012ultrasensitive,kristen2020amplitude,kakuyanagi2023submicrometer}

\vspace{5mm} 

A hybrid quantum sensor would then be a quantum system that harnesses a coupling to a different material system or a separate degree of freedom within the host material  
to gain a sensing advantage: 
either by enhancing the sensitivity by concentrating or amplifying the signal to be detected, 
or by transducing it to a separate degree of freedom where the sensor has a higher responsivity.

\vspace{5mm} 

It is worth noting that the term ``quantum sensing'' has been used in the literature in a variety of contexts;
for instance, a spin qubit operation in the context of quantum computing can also be considered a sensing operation insofar as it involves the measurement of a spin state.
Here, we focus on the more conventional idea of a sensor as a practical device that can be deployed to measure
a physical quantity in its external environment, such as electromagnetic fields, temperature, strain, and so on.

\vspace{5mm} 

From this perspective, 
we focus our discussion on hybrid quantum sensors based on NV centers in diamond 
as they are relatively mature, with commercial systems for NV scanning probe microscopy already starting to be available on the market~\cite{qnami,qzabre}.
The various diamond NV-based hybrid sensors are elaborated in Section 3.
Here, we highlight a few other notable examples of hybrid systems for quantum sensing.
For instance, 
a probe for external spins based on V2 spins in 4H-SiC coupled to a yttrium-iron garnet (YIG) nanostripe has been proposed and theoretically studied~\cite{tribollet2020hybrid}.
Quantum magnonics is another emerging field with potential sensing applications~\cite{awschalom2021quantum};
a recent work reported the sensing of single magnons are sensed with a quantum efficiency of 0.71 using a hybrid system consisting of a superconducting transmon qubit and a microwave cavity, 
where the cavity mediates the coupling between the superconducting qubit and Kittel mode on a YIG sphere via electric-dipole and magnetic-dipole couplings, respectively~\cite{lachance2020entanglement}.

\subsection{Criteria for Hybrid Systems} \label{SubSection:Metrics for Hybrid Systems}
How do we evaluate the functionality of a hybrid system? 
We propose here a set of criteria to evaluate the effectiveness of hybrid systems in the context of quantum computation, communication and sensing as discussed above. These criteria are intended as a way to categorize design considerations in building hybrid quantum systems. We survey various hybrid solid-state quantum systems with artificial atoms, and evaluate them according to these criteria in \textbf{Table \ref{tab:hyqusys}}. The purpose of this table is to serve as a guide to address the considerations and challenges in constructing hybrid quantum systems, as well as to present the state-of-the-art hybrid systems.

\vspace{5mm} 

\textbf{C1: Cooperativity.}
The cooperativity $C$ of a hybrid system with two coupled modes is given by the product of coupling-to-loss ratio of both modes \cite{clerk2020hybrid}:
$$C = \frac{g}{\kappa/2}\frac{g}{\gamma/2} = \frac{4g^2}{\kappa\gamma},$$
where $g$ is the coupling rate between the modes, and $\kappa$ and $\gamma$ are the loss rates of the respective modes. This is used as a measure of the degree of coherence of state transfer: $C$ greater than 1 indicates that coherent state transfer is possible between modes, as the coupling between modes occurs at a faster rate than the decoherence of either mode. In addition, a higher $C$ enables higher fidelity of state transfer in quantum communications~\cite{bersin2023telecom}. This metric is applicable to state transfer in quantum computing and communication, but is not relevant for quantum sensing.

\vspace{5mm} 

Among the hybrid quantum systems surveyed in Table \ref{tab:hyqusys}, many have cooperativities on the order of 100, signifying coherent state transfer between the different modalities. Of note are the large cooperativities on the order of 1000 for superconducting qubit coupled to phonons ($C \sim 1000$) \cite{chu2017quantum} and to photons ($C \sim 10^8$) \cite{yoshihara2017superconducting}. In the former case, the large cooperativity is made possible by operating in a dilution fridge, where the GHz-frequency superconducting qubit and phonon modes are in their quantum ground states without the need for further cooling. Additionally, by confining the phonon mode in a bulk crystal, phonon losses at interfaces and surfaces are avoided \cite{chu2017quantum}. In the latter case, the large cooperativity is enabled by the large coupling strength on the order of 100 MHz to almost 10 GHz \cite{yoshihara2017superconducting, scarlino2019coherent, landig2019virtual}. The large coupling strength of superconducting qubits arises from their large electric dipole moment, which is a factor of $\sim 10^4$ larger than the corresponding dipole moment of a single atom. \cite{blais2004cavity}

\vspace{5mm} 

\textbf{C2: Sensitivity.} 
The sensitivity $S$, applicable only for quantum sensing, can be defined as the minimum signal of a desired physical quantity that can be detected with a unit signal-to-noise ratio (SNR) over an integration time of one second~\cite{degen2017quantum}. This figure-of-merit is more relevant for quantum sensing than cooperativity, which is a dimensionless quantity that compares coupling rates to loss rates of two coupled systems.
In the literature, reports of sensor performance typically quote sensitivities in units of X/$\sqrt{\textrm{Hz}}$, where X is the unit of the relevant physical quantity (T for magnetic field, K for temperature, etc.).
Some published works which track the frequency shift of an optical resonance as a sensing mechanism would instead report a sensitivity in terms of a resonance shift of Hz/X.
As such, there is no obvious way to compare the performance across different types of sensors (e.g. magnetometers and thermometers).
Rather, it would be more meaningful to consider how the hybridization of different systems has enhanced the sensitivity of the particular hybrid quantum sensor, 
as well as how they perform compared to other state-of-the-art sensors for the relevant physical quantity.

\vspace{5mm} 

It can be shown that $S$ scales as~\cite{degen2017quantum}
$$S \sim \frac{1}{\xi \sqrt{T_c}},$$
where $\xi$ is the transduction parameter and $T_c$ is the sensor decoherence time.
It is desirable for a quantum sensor to have a large transduction parameter $\xi$ in order to efficiently convert the physical quantity to be sensed into detectable signal, as well as a long decoherence time $T_c$ to shield the sensor against environmental noise.
The examples covered in this work focus on the use of hybridization to enhance the physical signal, or to increase $\xi$ by transducing the signal to another degree of freedom with a higher sensitivity.

\vspace{5mm} 

\textbf{C3: Material Integration Considerations.}
For a hybrid system, it is also important to consider the ease of physical integration between different materials \cite{elshaari2020hybrid}. This is relevant because different components are more amenable to different material platforms, e.g. low-loss waveguides in silicon and silicon nitride, single-photon sources in III-V quantum dots or atomic-scale defects and dopants and superconducting nanowire single-photon detectors (SNSPDs) \cite{kim2020hybrid}. Combining multiple materials onto a single chip would make such a system more attractive to be deployed due to its smaller footprint and larger scalability, but can be challenging with the number of materials to be integrated. For this criteria, we take into account the compatibility of materials being integrated, and the type of integration technology used (e.g. wafer bonding, die bonding, transfer printing, pick-and-place).

\vspace{5mm} 

Across the hybrid systems surveyed, superconducting qubits and gate-defined spin qubits lend themselves to heterogeneous and flip-chip integration. Heterogeneous integration of multiple materials are required as superconducting materials (e.g. niobium (Nb) \cite{mi2017strong, mi2018coherent}, aluminum (Al) \cite{chu2017quantum, scarlino2019coherent, landig2019virtual}) that are often distinct from materials that are used to host gate-defined spin qubits (Si/SiGe heterostructure \cite{mi2017strong, mi2018coherent, borjans2020resonant}), photons (GaAs/AlGaAs heterostructure \cite{scarlino2019coherent, landig2019virtual}), and phonons (sapphire \cite{chu2017quantum}). On the other hand, the atomic size of spin qubits in defect and vacancy centers enable them to be integrated by embedding them in a host crystal through implantation or growth, followed by annealing or electron irradiation to form the desired centers \cite{wolfowicz2021quantum}. Integration of defect and vacancy center spin qubits with a separate target material can be achieved either through patterning their host crystal followed by transfer onto the target material, or by wafer bonding the host crystal to the target material followed by patterning the host crystal \cite{elshaari2020hybrid, kim2020hybrid}. 

\vspace{5mm} 

\textbf{C4: Readout Methods.}
In addition to integrating multiple components of a hybrid quantum system, the quantum state of its constituent modalities need to be read out. Certain modes have established sensors and detectors that can be used for readout. Examples include photon readout with avalanche photodiodes (APDs) or superconducting nanowire single photon detectors (SNSPDs) for photonic systems. However, certain modes are not amenable to direct detection either due to their nature (e.g. phonons) or their weak interaction (e.g. magnetic dipoles) \cite{rugar2004single}, and thus need to be converted to photons or charges that are easier to read out. For this criterion, we evaluate the methods available for reading out the quantum state of the constituents of the hybrid systems.

\vspace{5mm} 

\textbf{C5: Possibilities for Interconnects.}
To scale up hybrid quantum systems in the future, quantum states need to be transported between components via entanglement to accomplish various tasks in quantum communications, computing and sensing~\cite{diVincenzo2000thephysical,awschalom2021development}. These are accomplished by repeaters that transfer a quantum state over long distances, transducers that convert quantum state between different modalities, or frequency converters. For long-distance distribution, the use of optical photons is preferred due to their low losses, especially at telecommunications wavelengths in optical fibers where losses are typically as low as 0.2 dB/km \cite{awschalom2021development}, and mature fiber infrastructure is already present. Free-space optical transmission is also possible over larger wavelength ranges, with losses scaling only quadratically over distance compared to exponential losses in fibers \cite{awschalom2021development}. Microwave photonic links between superconducting qubit modules have also been demonstrated with low losses of 0.15 dB/km \cite{niu2023low}. However, such links require cryogenic cooling to millikelvin temperatures \cite{magnard2020microwave}, limiting their feasibility for long-distance distribution. As such, transduction between superconducting qubits and telecommunication-wavelength photon was demonstrated via phonons \cite{mirhosseini2020superconducting}, with single-photon to single-phonon-level signal transduction.

\begin{table}
\scriptsize
    \centering
    \RaggedRight
    \begin{tabularx}{\textwidth}{|X|X|X|X|X|X|} \hline 
        Hybrid system&  \multicolumn{5}{|c|}{Criteria}\\ \hline 
        &  C1: Cooperativity&  C2: Sensitivity&  C3: Material Integration Considerations&  C4: Readout Methods& C5: Possibilities for interconnects\\ \hline 
Spin qubit-photon&  
        \bb 100 (single SiV)~\cite{bhaskar2020experimental} \newline
        \bb 650 (Er ensemble)~\cite{wang2022high} \newline
        &
        \bb With MFCs\newline
        -- 195\,fT/$\sqrt{\textrm{Hz}}$~\cite{xie2021hybrid} (scalar)\newline
        -- 90\,pT/$\sqrt{\textrm{Hz}}$ \newline(vector)~\cite{wang2023hybrid}\newline
        -- 12\,pT/$\sqrt{\textrm{Hz}}$ (with $\mu$-diamonds)~\cite{shao2023high}\newline
        -- 0.57\,nT/$\sqrt{\textrm{Hz}}$ (with nanodiamonds)~\cite{chen2022nanodiamond}
        \newline \bb With MNPs\newline
        76\,$\mu$K/$\sqrt{\textrm{Hz}}$~\cite{liu2021ultra}
        \newline \bb Pressure-sensitive materials: Resonance shift \newline
        -- 36\,kHz/kPa~\cite{cai2014hybrid} (piezomagnetic)\newline
        -- 8.2\,kHz/kPa~\cite{kitagawa2023pressure} (magnetostrictive)
        \newline \bb Suspended diamond membrane: Resonance shift 2.3\,MHz/bar~\cite{momenzadeh2016thin}
        & 
        \bb color center embedded in diamond nanophotonic cavity~\cite{bhaskar2020experimental} \newline
        \bb embedded in fiber~\cite{chen2022nanodiamond} \newline
        \bb pick and place assembly~\cite{wan2020large} \newline
        \bb heterogeneous integration~\cite{borjans2020resonant,yu2023strong} &
        \bb via optical fiber~\cite{bhaskar2020experimental,wan2020large} \newline
        \bb SNSPD (in external setup)~\cite{bhaskar2020experimental,knaut2023entanglement} \newline
        \bb electrical readout for microwave photons~\cite{borjans2020resonant, yu2023strong}&
        \bb via optical fiber~\cite{bhaskar2020experimental} \newline
        \bb multiplexing~\cite{wan2020large, joshi2018frequency} \newline
        \bb entanglement with telecom photons~\cite{tchebotareva2019entanglement} or other nodes~\cite{pompili2021realization}\\ 
        \hline 
SC qubit-photon &
        \bb $2 \times 10^8$~\cite{yoshihara2017superconducting} &
        & 
        \bb heterogeneous: Al on Si~\cite{orgiazzi2016flux}
        &  
        \bb electrical~\cite{magnard2020microwave} \newline
        \bb photodiode~\cite{lecocq2021control} &
        \bb microwave link~\cite{magnard2020microwave, niu2023low} \newline
        \bb telecom photonic link~\cite{lecocq2021control} \newline
        \bb phonon transduction~\cite{mirhosseini2020superconducting}\\ 
        \hline 
SC qubit-phonon&
         \bb 1065~\cite{chu2017quantum}&
         &  
         \bb heterogeneous: Al on AlN on sapphire~\cite{chu2017quantum} \newline 
         \bb flip-chip bonding~\cite{dumur2021quantum} &
         \bb microwave~\cite{chu2017quantum} &
         \bb travelling phonon~\cite{dumur2021quantum} \\ \hline
gate-defined spin qubit-photon&
        \bb 38~\cite{mi2018coherent} \newline
        \bb 49~\cite{borjans2020resonant} \newline
        \bb 50 \cite{samkharadze2018strong}&
        & 
        \bb heterogeneous: Nb on Si/SiGe~\cite{mi2018coherent,borjans2020resonant} \newline
        \bb heterogeneous: NbTiN on Si/SiGe \cite{samkharadze2018strong} \newline
        \bb heterogeneous: NbTiN on GaAs/AlGaAs \cite{landig2018coherent}&
        \bb microwave~\cite{mi2018coherent,borjans2020resonant,samkharadze2018strong,landig2018coherent} &
        \bb electrical or microwave*\\ \hline
SC qubit-gate-defined spin qubit&
        \bb 214 (spin-photon), 40,248 (SC-photon)~\cite{landig2019virtual}
        &
        & 
        \bb heterogeneous: Al on GaAs/AlGaAs \cite{landig2019virtual}
        &
        \bb microwave using a SC qubit as readout~\cite{landig2019virtual}
        &
        \bb electrical or microwave*\\ \hline
SC qubit-spin qubit&
        \bb 644~\cite{pita2023direct}
        &
        & 
        \bb heterogeneous: QD in InAs/Al Josephson junction, embedded in SC circuit~\cite{pita2023direct}
        &
        \bb microwave~\cite{pita2023direct}
        &
        \bb electrical or microwave*\\ \hline
SC qubit-magnon-photon&
        \bb 749 (magnon-photon), 107 (SC qubit-magnon)~\cite{lachance2017resolving}
        &
        \bb Single magnon sensing with spherical ferrimagnetic crystal of YIG, efficiency\,=\,0.71~\cite{lachance2020entanglement} 
        & 
        \bb YIG sphere in 3D MW cavity~\cite{lachance2017resolving} - large footprint
        &
        \bb microwave~\cite{lachance2017resolving}
        &
        \bb microwave*\\ \hline
spin-photon-phonon&
        \bb 3 (spin-photon), $10^{-9}$ (spin-phonon)~\cite{shandilya2021optomechanical}
        &
        & 
        \bb NVs embedded in diamond microdisk cavity~\cite{shandilya2021optomechanical}
        &
        \bb optical fiber and free space~\cite{shandilya2021optomechanical}
        &
        \bb optical fiber~\cite{shandilya2021optomechanical}\\ \hline
spin-phonon&
        \bb 457~\cite{whiteley2019spin}
        &
        & 
        \bb color centers embedded in diamond or SiC~\cite{cai2014hybrid,kitagawa2023pressure,whiteley2019spin} \newline
        \bb heterogeneous: AlN on SiC~\cite{whiteley2019spin}
        &
        \bb optical readout of spin~\cite{cai2014hybrid,kitagawa2023pressure,whiteley2019spin}
        \bb electrical readout of phonon mode~\cite{whiteley2019spin}
        &
        \bb electrical or microwave*\\ \hline
    \end{tabularx}
    \caption{Benchmarking table comparing the metrics of the various demonstrated hybrid quantum systems involving artificial atoms discussed in this paper. Referenced performance levels are provided from reported literature cited in square brackets. 
    While some references provide quantitative values with corresponding uncertainties, 
    we have omitted the uncertainties in this Table to maintain consistency.
    Abbreviations used -- 
    MFC:~Magnetic flux concentrator;
    MNP:~Magnetic nanoparticle;
    SNSPD:~Superconducting nanowire single-photon detector;
    SC:~Superconducting;
    QD:~Quantum dot;
    MW:~Microwave.
    * Performance levels or demonstrations are not yet reported, but are based on the authors' suggestions.}
    \label{tab:hyqusys}
\end{table}

\section{Specific Hybrid Systems}

\subsection{Quantum Photonic Devices with Color Centers in Diamond} \label{Section:hybrid_cavity}

\subsubsection{Hybrid III-V Diamond Photonic Platform for Quantum Communications}
The combination of optical addressability with long spin coherence times makes color centers in diamond a promising candidate for repeater-based quantum networks. As discussed earlier in Section \ref{Section:QComm}, these networks could potentially find their applications in secure communication, modular quantum computing and distributed sensing. 

\vspace{5mm} 

In order to boost the entanglement generation rate in color-center-based quantum networks,  color centers can be integrated with nanophotonic devices. In particular,  coupling such artificial atoms to optical cavities could enhance atom-photon interaction and modify the emission properties of color centers, which improves spin readout and spin-photon entanglement fidelity. Recent advancements in diamond nanofabrication techniques have led to single-emitter cooperativities up to 105 \cite{bhaskar2020experimental}, along with the capabilities to interface with multiple color centers \cite{evans2018photon}. Additionally, nanophotonic devices can enable other functionality such as on-chip quantum frequency conversion, which is key to achieving long-distance quantum communications. Devices with low loss and small mode volume enable highly efficient nonlinear optical interactions at low pump powers, enhancing the signal-to-noise ratio at the single photon level\cite{li2016efficient}.

\vspace{5mm} 

\begin{figure}[thb]
\centering
\includegraphics[width=0.85\textwidth]{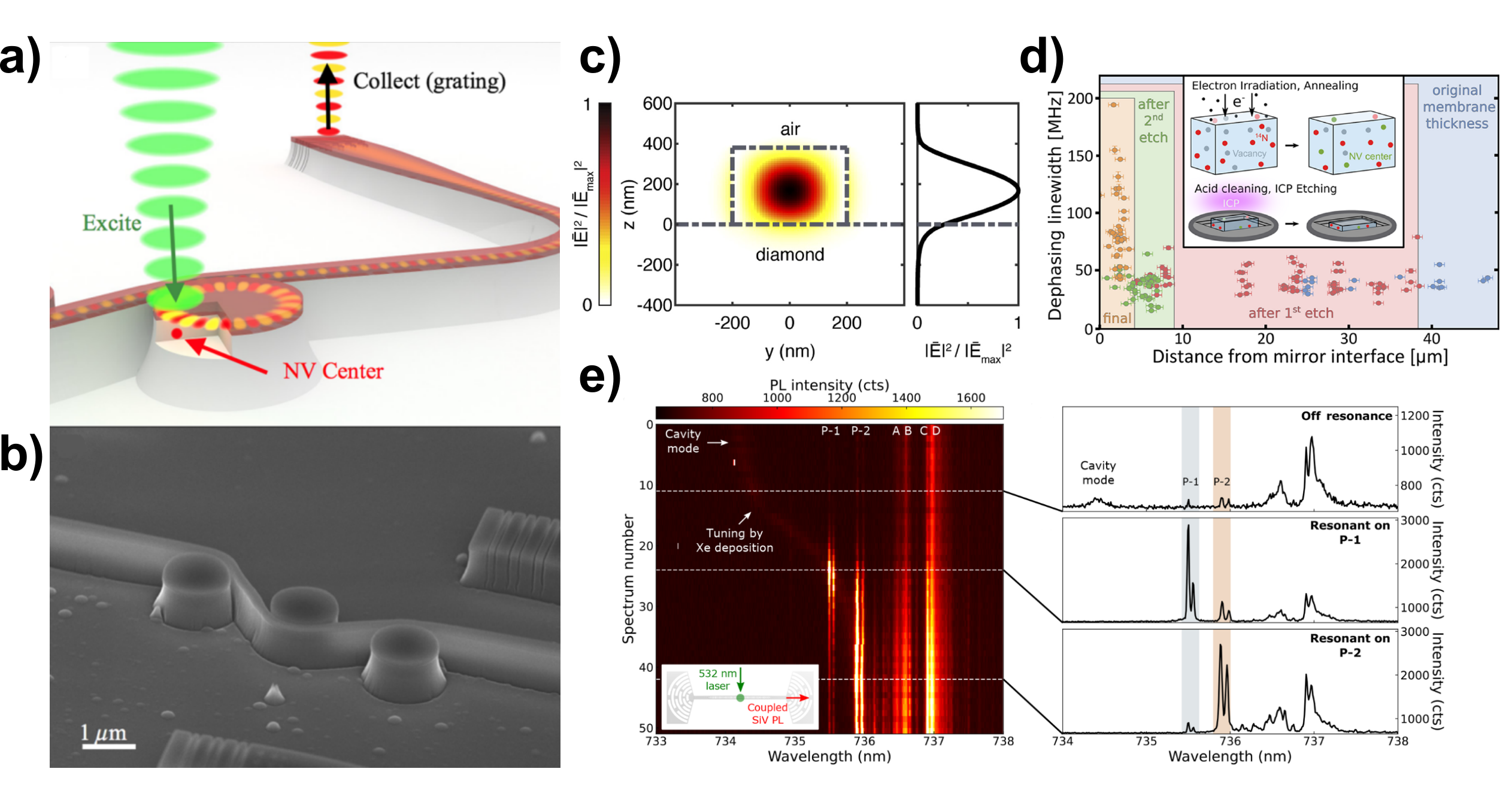}
\caption{Hybrid quantum photonic platform for Quantum Communications. (a) Schematic illustration of GaP-on-diamond platform for NV centers in diamond\cite{gould2016efficient}. Reprinted with permission\cite{gould2016efficient}. Copyright 2016, American Physical Society. (b) Scanning Electron Microscope image of the devices. Reprinted with permission\cite{gould2016efficient}. Copyright 2016, American Physical Society. (c) Normalized electric field distribution of GaP-on-diamond cavity resonant mode. This shows the evanescent field extends into the diamond and potential for coupling to color centers in diamond \cite{huang2021hybrid}. (d) Dephasing linewidth of NV centers at various distances from the diamond surface after fabrications. Reprinted under the terms of the Creative Commons Attribution Non-Commercial No Derivative Works 4.0 license (CC BY-NC)\cite{ruf2019optically}. Copyright 2019, American Chemical Society. (e) Emission enhancement of two SiV centers after tuning the GaP nanophotonic cavity on top of diamond into resonance. Reprinted with permission\cite{chakravarthi2023hybrid}. Copyright 2023, American Chemical Society.}
\label{fig:HybridIII-VPhotonics}
\end{figure}

Monolithic fabrication techniques of diamond nanophotonic devices involve milling or etching bulk single crystal diamond. These techniques are technically challenging and require precise control over the fabrication processes \cite{khanaliloo2015high, chia2022development}. More importantly, the process of etching diamond can have deleterious impacts on the properties of color centers in diamond (\textbf{Figure \ref{fig:HybridIII-VPhotonics}}(d)). The fabrication-induced surface states and strain can degrade the spectral stability of a single color center \cite{cui2015reduced, ruf2019optically, chakravarthi2021impact} or increase the spectral inhomogeneity across different color centers \cite{stas2022robust}, which are both undesirable for the quantum repeater protocols. The fabrication-induced damages have led to low device yield\cite{nguyen2019integrated} and precluded the on-chip integration of photonic components with other functionality, such as quantum frequency conversion and active devices. Additionally, there is currently no method for high-purity, wafer-scale growth of single crystal diamond\cite{nelz2019toward}, which constrains the scalability of this approach.

\vspace{5mm} 

In contrast, the development of nanofabrication techniques in III-V photonic materials is significantly more advanced. Nanophotonic components with diverse functionalities have been fabricated in III-V materials such as gallium arsenide (GaAs)\cite{dietrich2016gaas}, aluminum gallium arsenide (AlGaAs)\cite{baboux2023nonlinear} and gallium phosphide (GaP)\cite{wilson2020integrated}. Notably, GaAs photonic crystal cavities with quality factors -- a crucial performance metric for atom-cavity systems -- surpassing 100,000 have been used to couple to single quantum dots at near-infrared wavelengths\cite{kuruma2020surface}. State-of-the-art optical cavities on GaAs and AlGaAs have demonstrated quality factors exceeding 1 million\cite{kuruma2020surface, xie2020ultrahigh}, and those on GaP have reached over 100,000\cite{wilson2020integrated}, showcasing the potential of using III-V materials for low-loss, integrated quantum photonics platforms. In addition, recent progress on III-V quantum photonics has enabled key building blocks for repeater-based quantum communications such as single-photon router\cite{papon2019nanomechanical}, integrated photon detector\cite{reithmaier2015chip}, and efficient microwave-to-optical transduction\cite{stockill2022ultra, honl2022microwave}.

\vspace{5mm} 

A viable solution to the limitations of diamond nanofabrication is the heterogeneous integration of diamond with III-V photonic devices. Instead of directly fabricating devices on single-crystal diamond, the photonic device can be made in a high-index, III-V photonic layer on top of the diamond substrate. This allows optical modes to evanescently couple to color centers near the diamond surface. For NV centers in diamond, experimental demonstrations of this approach involves fabricating GaP waveguides or microdisks on diamond \cite{barclay2011hybrid, gould2016efficient, schmidgall2018frequency} (Figure \ref{fig:HybridIII-VPhotonics}(a-b)). A subsequent etching step into the diamond is required to form devices with high quality factors. Although it is possible to observe Purcell enhancement of optical emission, the etching step into diamond often degrades the spectral stability and homogeneity of the NV center optical transition\cite{chakravarthi2020inverse}. Recent advancements in integrating GaP photonic crystal cavities with negatively charged silicon vacancy (SiV) centers in diamond have shown significant promise in overcoming previous challenges (Figure \ref{fig:HybridIII-VPhotonics}(c)). By using a stamp-transfer technique, a cooperativity of 2 was successfully achieved in a hybrid device configuration \cite{chakravarthi2023hybrid} (Figure \ref{fig:HybridIII-VPhotonics}(e)). This was accomplished without the need for pre-selecting devices or aligning the SiV center with the photonic cavity, indicating the potential for further improvements of the atom-cavity coupling with hybrid systems. Additionally, efficient on-chip frequency conversions from the color center emission wavelengths to the telecommunication wavelengths with low-loss transmission in optical fibers are pursued on the hybrid device geometry \cite{huang2021hybrid, logan2023triply}. 

\subsubsection{Hybrid quantum sensors based on NV centers in diamond}
The most prominent application of NV centers in diamond is in magnetometry;
a change in the magnetic field shifts the energy levels of the NV center, 
which can be detected via an optically detected magnetic resonance (ODMR) technique.
Detailed explanations of NV magnetometry can be found in other in-depth reviews~\cite{abe2018tutorial,radtke2019nanoscale}.

\vspace{5mm} 

\begin{figure}[thb]
\centering
\includegraphics[width=0.85\textwidth]{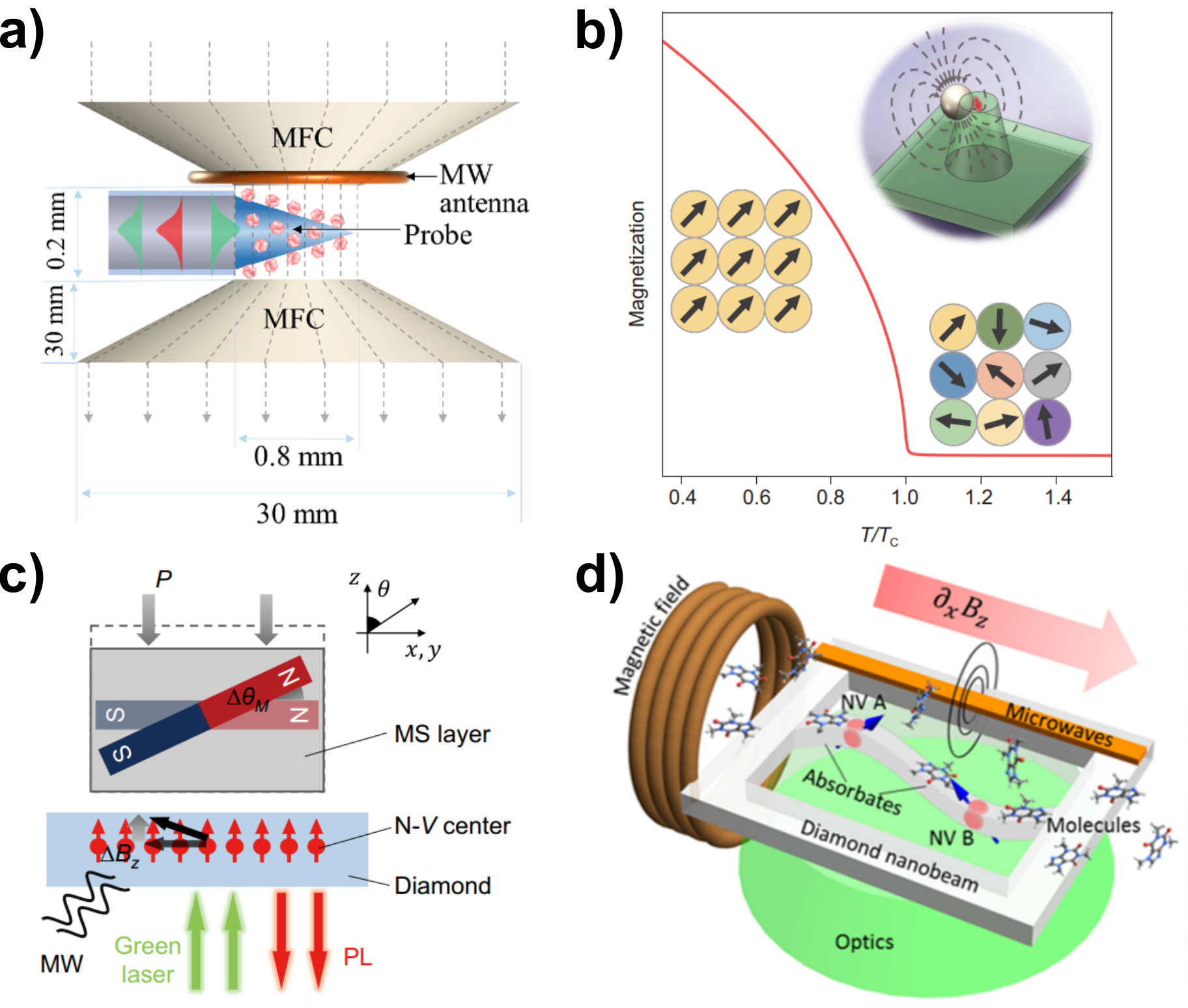}
\caption{Examples of hybrid quantum sensors based on nitrogen vacancy (NV) centers in diamond.
a) Magnetic field sensor: Nanodiamonds integrated with a tapered optical fiber, with magnetic flux concentrators (MFCs) enhancing the NV magnetic field sensitivity. Reprinted with permission from~\cite{chen2022nanodiamond}. Copyright 2022 American Chemical Society.
b) Nanothermometer: The critical magnetization of magnetic nanoparticles (MNPs) near the Curie temperature transduces a small temperature change to a large magnetic field signal measured by a nearby NV center~\cite{liu2021ultra}. 
Reprinted under the terms of the Creative Commons CC BY license.
c) Pressure sensor: A pressure applied to a magnetostrictive material alters the magnetic field sensed by the NV center. Reprinted with permission from~\cite{kitagawa2023pressure}. Copyright 2023 by the American Physical Society.
d) Mass sensor: In this proposed sensor, NV centers are coupled to the mechanical resonance of a nanobeam to detect the small masses of single molecules. Reprinted with permission from~\cite{barson2017nanomechanical}. Copyright 2017 American Chemical Society.
\label{fig:NV_sensor}}

\end{figure}

Here, we review hybrid quantum sensors based on NV centers, based on either bulk diamond or nanodiamond (ND) crystals,
featuring a hybrid coupling to a number of different systems: 

\vspace{5mm} 

\textbf{Magnetic Flux Concentrators (MFCs)}
NV centers in diamond membranes~\cite{fescenko2020diamond,xie2021hybrid,mao2023integrated,zhao2023pico,wang2023hybrid}
and nanodiamonds~\cite{chen2022nanodiamond} 
have been combined with high-permeability magnetic flux concentrators (MFCs) to enhance their magnetic field sensitivity
by collecting and focusing the magnetic field onto the sensor (see \textbf{Figure~\ref{fig:NV_sensor}}(a)).
With the MFCs capable of magnifying the field by a factor of $>$100, sensitivities for ensemble NV centers in diamond membranes have reached
195\,fT/$\sqrt{\textrm{Hz}}$~\cite{xie2021hybrid} 
and 90\,pT/$\sqrt{\textrm{Hz}}$~\cite{wang2023hybrid} for scalar and vector magnetometry, respectively.
For a tapered fiber with attached diamond micro- and nano-crystals, 
a sensitivity of 12\,pT/$\sqrt{\textrm{Hz}}$~\cite{shao2023high} and 
0.57\,nT/$\sqrt{\textrm{Hz}}$~\cite{chen2022nanodiamond} was achieved, respectively.

\vspace{5mm} 

\textbf{Magnetic Nanoparticles (MNPs)}
NV centers can operate as optical thermometers by transducing a small temperature change to a large change in the magnetic field
via a coupling to magnetic nanoparticles (MNPs)~\cite{zhang2018hybrid,wang2018magnetic,liu2021ultra}.
One strategy utilizes the critical magnetization near the Curie temperature, which can be designed by tuning the MNP chemical composition~\cite{wang2018magnetic} (see Figure~\ref{fig:NV_sensor}(b)).
With copper-nickel alloy MNPs, a sensitivity of 76\,$\mu$K/$\sqrt{\textrm{Hz}}$~\cite{liu2021ultra} was achieved near ambient conditions (38\,\degree C)~\cite{liu2021ultra}, 
which is orders of magnitude higher than the intrinsic $\sim$\,mK sensitivity of NV centers in bulk diamond~\cite{kucsko2013nanometre}.
Another report harnesses the volume phase transition in a hydrogel to induce a sharp variation in the separation distance between the MNP and NV centers in ND,
achieving a sensitivity of 
96\,mK/$\sqrt{\textrm{Hz}}$~\cite{zhang2018hybrid},
an order of magnitude higher than a bare ND.

\vspace{5mm} 

\textbf{Pressure-sensitive Magnetic Materials}
Materials with magneto-mechanical properties can be utilized to transduce a pressure or stress signal to a change in the magnetic field.
By incorporating a NV center coupled to a piezomagnetic layer, the pressure signal is amplified by three orders of magnitude compared to its direct effect on the diamond, 
with an observed NV resonance frequency shift of up to 36\,kHz/kPa~\cite{cai2014hybrid}.
The authors inferred an optimal achievable sensitivity of $\sim$\,0.35\,kPa/$\sqrt{\textrm{Hz}}$ and
$\sim$\,75\,fN/$\sqrt{\textrm{Hz}}$ for stress and force, respectively.
In another report, pressure applied to a magnetostrictive material 
rotates its magnetization, resulting in a resonance frequency shift of $8.2\pm0.9$\,kHz/kP
~\cite{kitagawa2023pressure} (see Figure~\ref{fig:NV_sensor}(c)). 

\vspace{5mm} 

\textbf{Nanomechanical Systems}
NV centers can be used to sense the motional degree of freedom in nanomechanical systems, such as cantilevers made of diamond~\cite{ovartchaiyapong2014dynamic} or glass~\cite{dadhich2023nanodiamonds}.
A suspended diamond membrane hosting NV centers has been shown to detect a displacement due to an applied pressure of $\approx$\,40\,Pa, 
as well as an associated resonance frequency shift of $\approx$\,2.3\,MHz/bar~\cite{momenzadeh2016thin}.
A proposed sensor based on a diamond nanobeam is predicted to achieve a mass sensitivity of $\approx$\,1\,zg/$\sqrt{\textrm{Hz}}$, 
equivalent to 50 carbon atoms~\cite{barson2017nanomechanical}  (see Figure~\ref{fig:NV_sensor}(d)).

\vspace{5mm} 

While NV centers are well-known quantum sensors in their own right,
their applicability in hybrid systems can also be attributed to their relative ease of integration with other material systems:
diamond is a stable and inert host material which often does not require sophisticated techniques to interface with a different material system. 
It is commercially available in both bulk diamond and nanodiamond forms,
and be operated conveniently in ambient conditions without needing cryogenic environments.
Sensor control and signal readout can be readily performed optically via optical fibers.

\vspace{5mm} 

These favorable properties have motivated many efforts in developing NV-based quantum sensors as practical devices that can be effectively deployed in real-world settings, where much emphasis is placed on the integration of multiple functional components (e.g. power supplies, optical fiber coupling, control electronics, etc.) together with the NV-hosting diamond into a single compact or portable package. For instance, NV centers in mm-scale diamond plates have been integrated with optical fibers and microwave circuit in a compact package~\cite{kuwahata2020magnetometer,sturner2021integrated}, 
while other efforts incorporated nanodiamond crystals hosting NV centers onto or directly within optical fibers~\cite{liu2013fiber,maayani2019distributed,bai2020fluorescent,filipkowski2022magnetically,chen2022nanodiamond}.
We can be optimistic that more integration efforts will follow for NV sensors operating with these hybrid modalities,
enabling their use in practical quantum sensing applications.

\subsection{Spin-Phonon Systems}
Quantum computing needs qubits to interact with each other so as to perform gate or entangling operations, as well as a quantum memory to store quantum information if not being processed \cite{perez2011quantum}. Current mature quantum computing platforms include trapped ions and superconducting circuits. Trapped ion quantum computing involves the use of electric and/or magnetic fields to trap ions at specific positions, and the qubit state is encoded in the electronic levels of the ions. Individual qubit states are manipulated using lasers, while multi-qubit interactions are mediated by common motional modes \cite{cirac1995quantum, bruzewicz2019trapped, brown2021materials}. Superconducting circuit quantum computing utilizes microwave circuits made  from superconducting materials, and the qubit state is encoded in the anharmonic energy levels arising from nonlinear elements included in the circuit. Single-qubit gates are implemented by applying microwave pulses to each qubit, while multi-qubit gates are implemented through circuits that couple individual qubits \cite{kjaergaard2020superconducting, blais2020quantum, blais2021circuit}.

\vspace{5mm} 

While mature, both platforms face different challenges with respect to scaling up. In trapped ion quantum computing, scaling up to more trapped ion qubits increases the separation between distant qubits, which increases the two-qubit gate time \cite{wright2019benchmarking}. For superconducting circuits, noise due to microwave crosstalk increases as more superconducting qubits are introduced, in addition to the increased footprint and thermal load \cite{blais2020quantum}. Multi-qubit gates are often limited to physical neighbors in planar architectures \cite{gao2021practical}, in contrast with trapped ions where non-neighboring qubits can be entangled through motional modes.

\vspace{5mm} 

As such, an alternative platform being pursued is defect spins in solid-state coupled to phonons in nanomechanical resonators. Qubit states can be encoded in the spin angular momentum states of such defect spins. By virtue of being embedded in solid state, no additional overhead is required to trap these spin qubits unlike trapped ions. Multiple spins can also be embedded in the same phonon mode. Their atomistic size scales also makes them more amenable to scaling up as compared to superconducting circuits, as more qubits can be packed in the same area.

\vspace{5mm} 

In coupling to spins, photons have thus far been more popular due to the existence of optically-accessible spin defects in which different spin states can be addressed through different optical transitions, and the ability to transmit photons over long distances \cite{atature2018material, wolfowicz2021quantum}. However, there is a mismatch even within this modality: while photons can be transmitted in optical fibers with lowest losses at telecommunications wavelengths, optical transitions associated with spin defects are often in the visible or near-infrared, which have higher transmission losses. Additionally, the intrinsic coupling between the electric dipole of a spin defect and a single photon is weak, and requires optical cavities with small mode volumes to enhance the coupling \cite{atature2018material}.

\vspace{5mm} 

Therefore, phonons in nanomechanical resonators have emerged as a candidate to mediate interactions with spins \cite{treutlein2014hybrid, lee2017topical}. An attractive quality is the lower footprint offered by nanomechanical resonators: as the speed of sound in solid is 4-5 orders of magnitude slower than that of light in the same solid, the characteristic wavelength of a phonon is correspondingly smaller than that of a photon at the same frequency. As a result, phononic devices can be miniaturized relative to their photonic counterparts \cite{safavi2019controlling}. In particular, spin transitions are often at microwave frequencies (few GHz), and interfacing these with nanomechanical resonators would allow for resonant interactions in smaller footprints compared to superconducting circuits operating at similar frequencies.

\vspace{5mm} 

The spin-phonon interaction is analogous to spin-photon interaction, in that both phonons and photons are bosons; thus, the interaction Hamiltonians are similar. The spin-phonon (photon) coupling rate is proportional to acoustic (electric) dipole moment and zero-point phonon (photon amplitudes). In order to achieve a strong spin-phonon coupling rate, it is desirable to choose a spin that is very sensitive to strain fields in phonons, and embed it in a structure that has large zero-point phonon amplitudes.

\vspace{5mm} 

Current implementations have used the nitrogen vacancy (NV) spin in diamond embedded in bulk resonators \cite{macquarrie2013mechanical, macquarrie2015coherent, chen2018orbital}, cantilevers \cite{teissier2014strain, ovartchaiyapong2014dynamic, lee2016strain, meesala2016enhanced} or microdisk resonators \cite{shandilya2021optomechanical}. However, the single-phonon cooperativities are limited by the low strain susceptibility of the ground state orbitals ($\sim$ 10 GHz/strain for NV) \cite{teissier2014strain}. 
The aforementioned resonators feature relatively large mode volumes due to the lack of confinement. Surface acoustic wave (SAW) resonators offer confinement in the normal direction and thus smaller mode volumes and larger zero-point phonon amplitudes. SAW resonators have been used to interface with spin qubits, including NV centers \cite{golter2016optomechanical, golter2016coupling}, silicon vacancy (SiV) centers \cite{maity2020coherent} and nuclear spins in diamond \cite{maity2022mechanical}, as well as color centers in silicon carbide \cite{whiteley2019spin, hernandez2020anisotropic, hernandez2021acoustically}. 

\vspace{5mm} 

\begin{figure}[thb]
\centering
\includegraphics[width=0.6\textwidth]{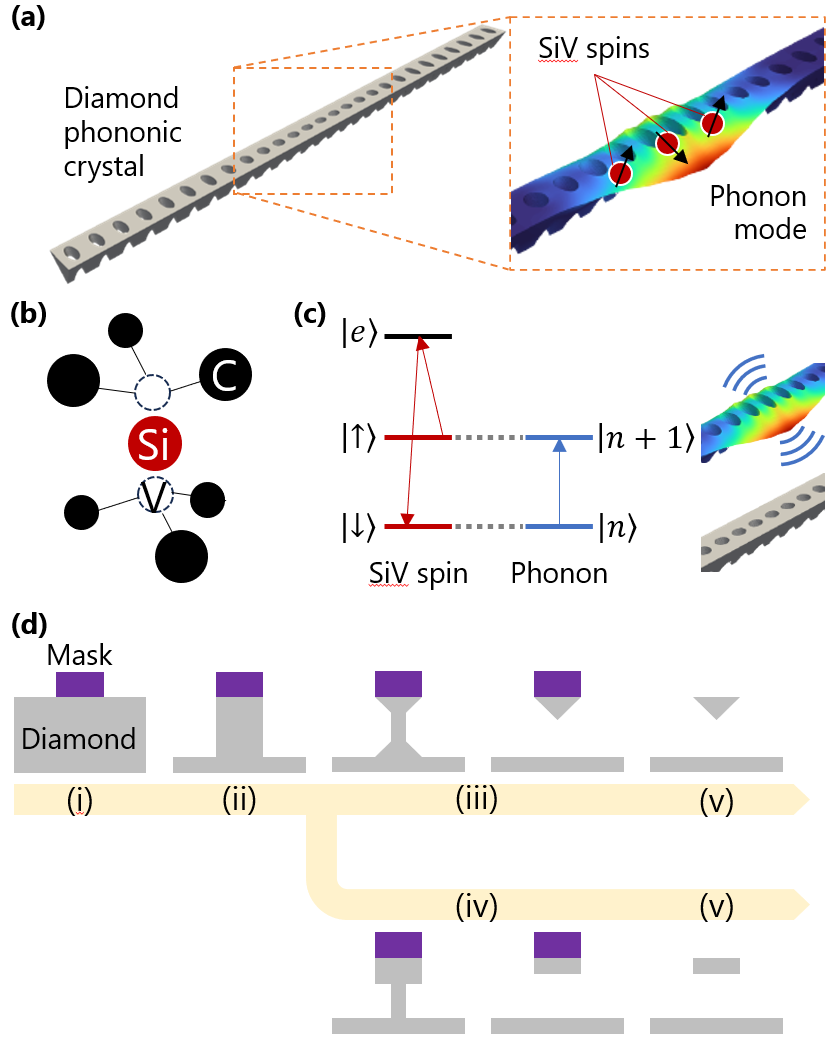}
\caption{Proposed spin-phonon system with silicon vacancy (SiV) centers in diamond phononic crystal. (a) Schematic of diamond phononic crystal with embedded SiV spins interacting with phonon mode. (b) Structure of SiV center in diamond (C: carbon). (c) Relevant energy levels in spin-phonon system with spin qubit levels $\{ |\uparrow \rangle, |\downarrow \rangle \}$ and phonon mode occupancy $n$. $|e \rangle$ denotes excited state. (d) Fabrication of suspended nanomechanical resonators from bulk diamond substrates via (i) mask definition, (ii) vertical etch, (iii) angled ion beam etching, (iv) quasi-isotropic etching and (v) mask removal to reveal completed triangular or rectangular cross-section structures.
\label{fig:spin-phonon}}

\end{figure}

A proposal to implement spin-phonon quantum computing would comprise of defect spins embedded in a suspended phononic crystal \cite{meesala2018strain}, in which qubits are defined in the defect spins and localized phonon modes in the phononic crystal mediates inter-qubit entanglement. (\textbf{Figure \ref{fig:spin-phonon}}(a)) The SiV center in diamond, comprising of an interstitial silicon atom between two vacancies in the diamond lattice (Figure \ref{fig:spin-phonon}(b)), has been proposed as a candidate for strong spin-phonon coupling given its ground-state strain susceptibility of 1 PHz/strain that is 5 orders of magnitude larger than that of the NV center ground-state orbitals \cite{meesala2018strain, shandilya2021optomechanical}. Additionally, using the ground state precludes spontaneous emission that is typical of excited state transitions used in current implementations \cite{whiteley2019spin, golter2016optomechanical}. 

\vspace{5mm} 

Embedding it in a phononic crystal, a quasi-periodically patterned structure that confines a desired phonon mode in all three dimensions, would produce large zero-point phonon amplitudes (Figure \ref{fig:spin-phonon}(c)). In contrast with photons, which can leak into free space, engineered phonon modes in phononic crystals that are isolated from the substrate cannot leak out of the structure \cite{safavi2019controlling}. As such, phonon modes have higher quality factors than photon modes at the same frequency. Fabricating such phononic crystals in diamond, with its low thermoelastic damping and high Young's modulus \cite{tao2014single}, would enable phonon modes to be realized with low damping rates compared to other materials.

\vspace{5mm} 

To realize such structures with minimal phonon damping losses, it is ideal to create suspended structures. However, diamond lacks a mature heteroepitaxial platform like silicon-on-insulator. As such, bulk micromachining techniques have been developed to create structures suspended from the substrate. Such techniques include angled ion beam etching \cite{chia2022development} and quasi-isotropic etching \cite{khanaliloo2015high} (Figure \ref{fig:spin-phonon}(d)). However, surface roughness is a limiting factor due to mask erosion in the former \cite{chia2022development} and uneven etching in the latter \cite{wan2018two}. Alternatively, thin-film diamond membranes have emerged as a promising platform, where recent progress has been made in optimizing the smoothness of such membranes. These membranes are transferred using stamp transfer printing over trenches defined in a carrier wafer, allowing for suspended structures to be defined in the transferred membranes \cite{guo2021tunable}. SiV centers are introduced at targeted locations through focused ion beam implantation \cite{schroder2017scalable, sipahigil2016integrated} or ion beam implantation through masks \cite{toyli2010chip, nguyen2019integrated}, with lateral accuracy limited to that of electron beam lithography used to define the mask \cite{nguyen2019integrated}.

\vspace{5mm} 

Single qubit operations can be performed through optical addressing, while multi-qubit operations are mediated via the phonon mode \cite{lemonde2018phonon}, analogous to trapped ion quantum computing. Flipping of a qubit state is accompanied by the absorption or emission of a phonon in the phonon mode (Figure \ref{fig:spin-phonon}(c)) The advantage of such a system lies in the low-loss nature of the phonon mode confined in the suspended phononic crystal, making interactions between vastly-separated spins more viable than in trapped ions. Using phonon modes at GHz frequencies resonant with spin transitions also allow for faster gate operations as compared to motional modes in trapped ions, which are on the order of MHz. Phonons also intrinsically couple to a variety of modalities, such as photons and electric fields, as well as other artificial atoms \cite{lee2017topical}. Such interconnects can be realized through phonon waveguides supporting traveling phonon wavepackets \cite{fang2016optical}, or by heterogeneous integration of piezoelectric thin-films on non-piezoelectric substrates, such as diamond \cite{maity2020coherent}, silicon \cite{mirhosseini2020superconducting} and silicon carbide \cite{whiteley2019spin}. As such, phonons also provide a way to interconnect multiple modules for scaling up of hybrid quantum systems.

\vspace{5mm} 

\vspace{5mm} 

\subsection{Hybrid Superconducting Circuits with van der Waals Materials} \label{sec:sc-vdw}

Quantum computation using artificial atoms made out of Josephson-junction-based, superconducting circuits has enabled the demonstrations of small-scale quantum algorithms, such as break-even quantum error correction \cite{ofek2016extending, sivak2023real, ni2023beating}, quantum machine learning \cite{havlivcek2019supervised, harrigan2021quantum}, calculations of energies of small molecules \cite{o2016scalable, google2020hartree}, and quantum simulations of Bose-Hubbard model \cite{yan2019strongly, ma2019dissipatively}. The appeal of this platform lies in its relatively long coherence times, high-fidelity gate operations, and the adaptability in design parameters, making it a promising candidate for implementations of quantum information processing tasks as described in Section \ref{Section:QC}. This type of artificial atom can be engineered to interact strongly with microwave photons \cite{blais2021circuit}. The interaction has led to atom-cavity system in the ultrastrong coupling regime with high cooperativities \cite{yoshihara2017superconducting}. In addition, the same mechanism results in high fidelity readout of the qubit state and connections between superconducting qubits over a longer distance \cite{magnard2020microwave,niu2023low}. It's worth noting, in addition to microwave photons, a recent demonstration showed that optical photons can be used to control and read out the superconducting qubit state \cite{lecocq2021control}. This presents a promising pathway in tackling the bottleneck in wiring of large-scale superconducting-qubit-based quantum processors.

\vspace{5mm} 

High-quality superconducting qubits have stringent requirements on the cleanliness of materials surfaces and interfaces \cite{de2021materials}. In particular, the Josephson tunnel junction, which consists of two superconductors separated by a thin barrier through which the Cooper pairs can tunnel coherently, could house several sources of noise and dissipation. The prevalent use of aluminum oxide as the tunneling barrier in modern superconducting qubit architectures has been identified as a limiting factor. Given its amorphous nature, the aluminum oxide contains a high concentration of defects. When the superconducting qubit couples to a discrete defect in the amorphous oxide, the qubit's energy relaxation time, $T_1$, is shortened. Towards mitigating the defects associated with amorphous oxide and interfaces, van der Waals (vdW) heterostructures have been considered for constructing the Josephson junctions. The family of vdW materials contains high-quality superconductors and crystalline dielectrics. When assembling via an entirely dry process, the vdW heterostructures are posited to yield atomically pristine interfaces which can reduce the sources of loss for superconducting qubits \cite{siddiqi2021engineering,liu20192d}.

\vspace{5mm} 

\begin{figure}[thb]
\centering
\includegraphics[width=0.85\textwidth]{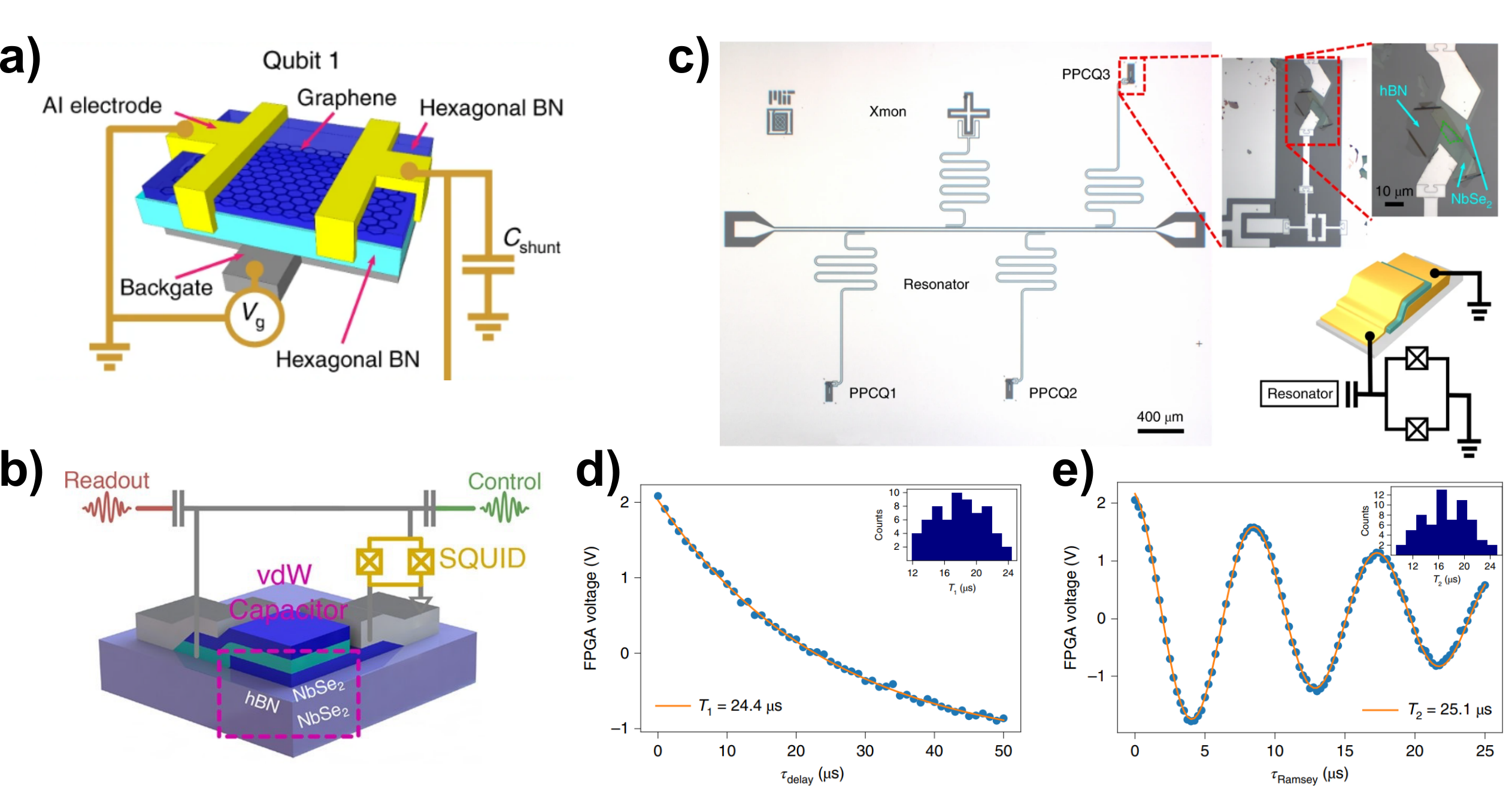}
\caption{Novel superconducting qubits with vdW materials. (a) Schematic illustration of using voltage-tunable van der Waals heterostructures for superconducting gatemon. Reprinted with permission\cite{wang2019coherent}. Copyright 2019, Springer Nature. (b) Schematic illustration of replacing the coplanar capacitor of a superconducting transmon qubit with a parallel-plate capacitor made out of vdW heterostructures. Reprinted with permission\cite{antony2021miniaturizing}. Copyright 2022, American Chemical Society. (c) Optical image of a transmon qubit shunted by a vdW parallel-plate capacitor, showing the contrast between the size of conventional transmon qubit with the hybrid transmon qubit. Reprinted with permission\cite{wang2022hexagonal}. Copyright 2022, Springer Nature. The hybrid superconducting qubit shows a $T_1$ of 24.4 $\mu$s (d) and $T_2$ of 25.1 $\mu$s (e). Reprinted with permission\cite{wang2022hexagonal}. Copyright 2022, Springer Nature.} 
\label{fig:vdW_SCqubit}
\end{figure}

There is a wide variety of superconducting qubits with differing configurations of circuit elements. A relatively new type of superconducting qubits, gatemon, uses a voltage-tunable semiconductor weak link as the tunnel barrier for the Josephson junction. By employing ballistic transport in high mobility systems, this design can overcome the dissipative issues associated with diffusive normal metal barriers. The gatemon was first demonstrated using Indium Arsenide (InAs) nanowires and two-dimensional electron gas in InAs/InGaAs heterostructures \cite{larsen2015semiconductor,luthi2018evolution,casparis2018superconducting}. Recently, a hybrid gatemon made out of superconducting aluminum and graphene was demonstrated to have a $T_1$ of tens of nanoseconds and a dephasing time, $T_2^*$, of $\sim$ 55 ns \cite{wang2019coherent} (\textbf{Figure \ref{fig:vdW_SCqubit}}(a)).

\vspace{5mm} 

Additional works along this direction used capacitors made with vdW heterostructures to build superconducting transmon qubits \cite{wang2022hexagonal, antony2021miniaturizing}. The vdW heterostructure, consisting of a vdW dielectric layer sandwiching in between two vdW superconductor layers, can be considered as a compact, parallel-plate capacitor (Figure \ref{fig:vdW_SCqubit}(b)). The large coplanar capacitors used by superconducting transmon qubits were adopted to reduce the sensitivity of such qubits to defects at the materials surfaces and interfaces. By using the parallel-plate capacitors made with vdW heterostructures as the shunting capacitors,  the hybrid superconducting transmon qubits demonstrated coherence times on par with conventional superconducting qubits (Figure \ref{fig:vdW_SCqubit}(d-e)). Notably, this was achieved with a footprint that is 250 times smaller than the typical superconducting qubits (Figure \ref{fig:vdW_SCqubit}(c)). 

\vspace{5mm} 

Although vdW heterostructures have been extensively characterized through DC transport measurements and optical spectroscopy, their responses in the microwave frequency regime remains relatively unexplored. This gap opens substantial prospects for incorporating vdW materials into superconducting quantum circuits. Beyond the advantages of mitigating known decoherence channels, the application of vdW materials significantly enhances the flexibility in the design of hybrid quantum systems. Conventional approaches for integrating materials into superconducting quantum circuits are often hindered by the lattice-matching constraints of direct growth techniques, or the availability of metal sputtering and evaporation processes. In contrast, the assembly of vdW heterostructures relies on a straightforward mechanical stacking method. This approach has been proven to be effective in integrating a diverse range of materials without the constraints of lattice mismatch and other material incompatibilities. The incorporation of vdW heterostructures into superconducting quantum circuits not only broadens the scope of material combinations but also presents new possibilities for quantum circuit designs, such as the introduction of novel methods for encoding information \cite{siddiqi2021engineering}.

\vspace{5mm} 

\subsection{Hybrid Superconductor-Semiconductor Quantum Systems with Spin Qubits}

Gate-defined spin qubits in semiconductors is another promising hardware platform in building solid-state quantum processors. Here, an artificial atom is defined by the electrostatic potential landscapes, which are shaped by the electrodes. This enables the confinement of electrons or holes within the semiconductor through these electrostatic potentials. The spin degree of freedom of isolated charge carriers can then be exploited to serve as qubits. This type of solid-state qubit has demonstrated long coherence times and high-fidelity gate operations. State-of-the-art experiments in this field have achieved $T_1$ on the order of seconds \cite{amasha2008electrical,prance2012single,yang2013spin,borjans2019single,camenzind2018hyperfine,chan2018assessment,ciriano2021spin,hofmann2017anisotropy,hollmann2020large}. In addition, significant progress have been made towards pushing the fidelity for single- and two-qubit gate operations \cite{burkard2023semiconductor,stano2022review}, with single-qubit gate fidelity exceeding 99.9$\%$ \cite{yoneda2018quantum, yang2019silicon} and two-qubit gate fidelity surpassing 99$\%$ \cite{mills2022two,xue2022quantum,noiri2022fast,weinstein2023universal}.

\vspace{5mm} 

Most of the two-qubit quantum logic operations between gate-defined spin qubits predominantly rely on nearest-neighbor interactions. The interactions arise from the wavefunction overlaps of two charges that are confined at close proximity to each other. While this scheme can be implemented straightforwardly and had demonstrated fast gate operation times, it presents significant complexities for the design, fabrication, and operation of large-scale quantum processors. In order to employ this scheme of two-qubit gate operations, adjacent qubits can only be separated by 100 - 200 nm \cite{burkard2023semiconductor,petta2005coherent}. Integration of a large number of qubits at this scale presents formidable challenges in device yields and fan-out of control and readout wires. 

\vspace{5mm} 

\begin{figure}[thb]
\centering
\includegraphics[width=0.85\textwidth]{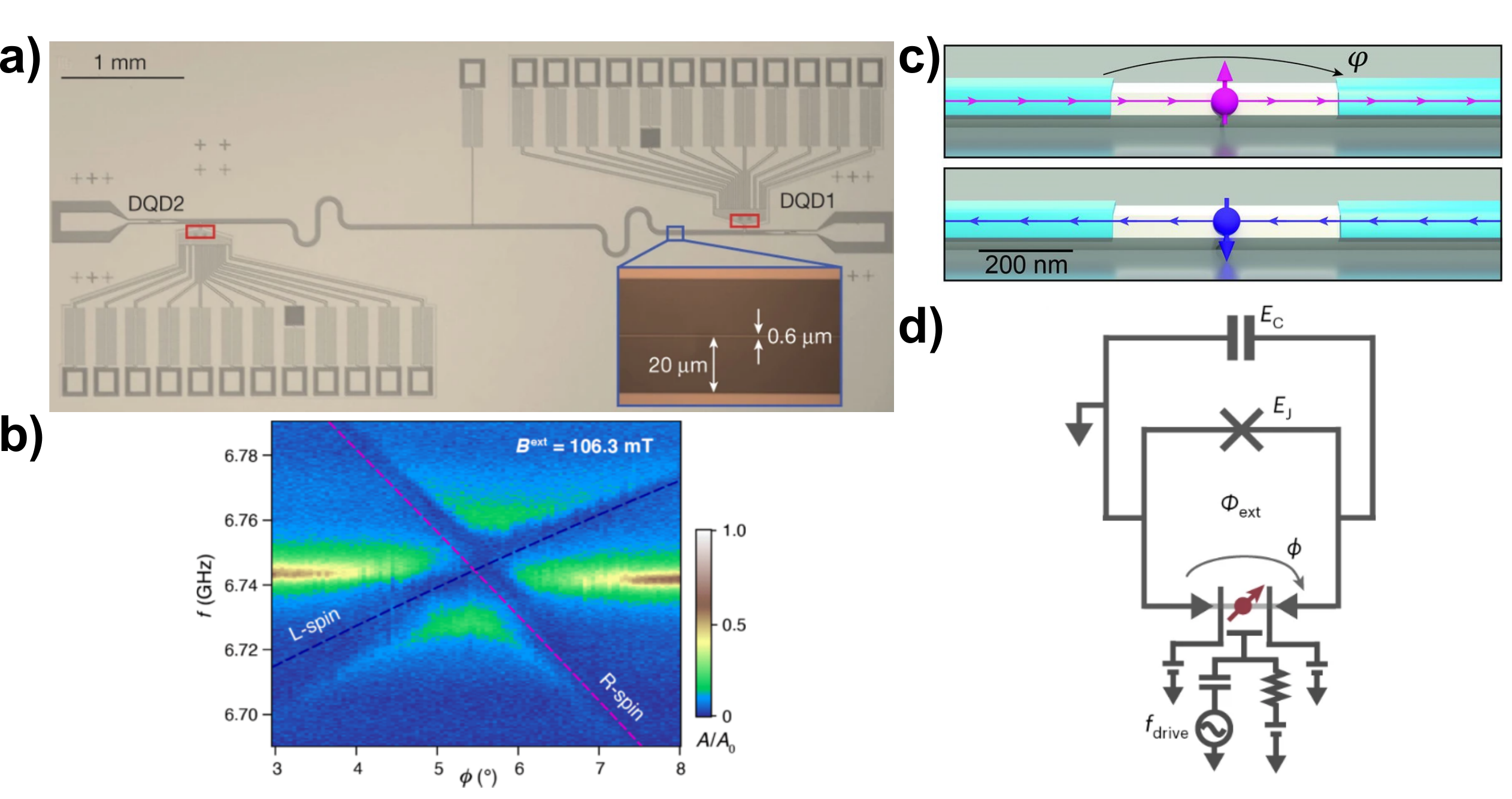}
\caption{Hybrid superconductor-semiconductor quantum systems. (a) Optical microscope image of a superconducting microwave resonator connecting two spatially separated double quantum dots. Reprinted with permission\cite{mi2018coherent}. Copyright 2018, Springer Nature. (b) Resonant spin-photon-spin coupling. The avoided crossings, a signature of strong spin-photon coupling, are observed when both gate-defined spin qubits are tuned into resonance with the cavity. Reprinted with permission\cite{borjans2020resonant}. Copyright 2020, Springer Nature. (c) Schematic illustration of Andreev spin qubits. The supercurrent across the Josephson junction is spin-dependent, allowing the spin qubit to be coupled to superconducting quantum circuits. Reprinted with permission\cite{hays2021coherent}. Copyright 2021, The American Association for the Advancement of Science. (d) Embedding the Andreev spin qubit into a superconducting transmon qubit provides means to readout and manipulate the Andreev spin qubits. Reprinted with permission\cite{pita2023direct}. Copyright 2023, Springer Nature.} 
\label{fig:SuperSemi_SpinQubit}
\end{figure}

Hybrid quantum systems with superconducting cavities integrated with semiconductor spin qubits show promising results for realizing long-distance qubit interactions \cite{burkard2020superconductor} (Figure \ref{fig:SuperSemi_SpinQubit}(a)). These systems leverage microwave cavities to couple two spatially separated spin qubits. The microwave photons within these cavities can mediate coherent interactions between the qubits.  Pioneering experiments in this direction include the demonstrations of strong coupling between the microwave photon and the spin qubit \cite{mi2018coherent,samkharadze2018strong,landig2018coherent} and cavity-mediated interactions between spin qubits \cite{borjans2020resonant,harvey2022coherent} (Figure \ref{fig:SuperSemi_SpinQubit}(b)). This progress has eventually led to the recent milestone of two-qubit gate operations between spin qubits separated by hundreds of micrometers \cite{dijkema2023two}. The hybrid superconductor-semiconductor device architecture offers a promising pathway to connect spatially separated modules of spin qubits together, potentially solving the scalability bottleneck in large-scale quantum processors based on spin qubits. 

\vspace{5mm} 

The hybrid integration of superconductor-semiconductor systems presents a novel pathway for linking gate-defined spin qubits with other quantum information processing platforms, such as superconducting qubits. As discussed in Section \ref{Section:QC}, this approach can be used to build quantum processors that leverage on the unique advantages of each qubit type. In a way similar to the cavity-mediated coupling between spin qubits, early work in this area used virtual photons in a microwave resonator to mediate interactions between a gate-defined spin qubit and a superconducting qubit \cite{landig2019virtual}. More recently, the development of Andreev spin qubits \cite{chtchelkatchev2003andreev,padurariu2010theoretical} presents exciting opportunities (Figure \ref{fig:SuperSemi_SpinQubit}(c)). The Andreev spin qubit exploits the spin degree of freedom of a quasiparticle occupying an Andreev bound state in a Josephson junction \cite{hays2021coherent}. The integration of an Andreev spin qubit directly into a superconducting qubit has led to demonstrations of coherent couplings between the two \cite{pita2023direct,bargerbos2022singlet,bargerbos2023spectroscopy} (Figure \ref{fig:SuperSemi_SpinQubit}(d)). While these developments are promising, much work remains to fully understand and capitalize on the potential advantages of such hybrid quantum systems over traditional monolithic quantum computing architectures. 

\section{Conclusion and Outlook}
In conclusion, we have conducted a comprehensive survey of hybrid solid-state quantum systems based on artificial atoms and attempted to provide a perspective on this area of development based on the type of hybrid system, and a proposed set of criteria. We first established the set of criteria based on practically desirable characteristics of hybrid quantum systems: (1) cooperativity, (2) sensitivity, (3) material integration, (4) readout methods, and (5) possibility of interconnects.  Thereafter, we used this to characterize various hybrid solid-state quantum systems ranging from artificial atoms interacting with particles or quasiparticle systems to hybrid systems comprising different artificial atoms. Table \ref{tab:hyqusys} provides a summary of the performance and challenges/opportunities that each hybrid system offers for each of the practical criteria discussed. We find that hybrid quantum systems coupling spin qubits to photons represent a heavily explored option with demonstrated outcomes for all criteria compared to other hybrid manifestations. However, some hybrid options have already begun to show significant advantages in certain criteria. For example, the superconducting qubit and photon pairing excels in cooperativity and has proven readout methodologies and feasible interconnection options. Others, like the gate-defined spin qubit and photon pairing offers a range of material options for integration and have proven read-out techniques, but have yet to show outstanding cooperativity. In the arena of quantum sensing, mainly spin qubit-photon hybrid and the superconducting qubit-magnon-photon hybrid have demonstrated performance. It is foreseeable that more examples for each type of hybrid system will be reported and new hybrid variants invented.

\vspace{5mm} 

Although we have covered mainly hybrid quantum systems that have been experimentally demonstrated, there are likely many more such systems in ideation stages or being proposed in theoretical papers but yet to be realized. Some new material systems may hold promise for hybrid quantum system implementations. We have already mentioned van der Waals heterostructures being used in conjunction with superconducting transmon qubits in Section \ref{sec:sc-vdw}. Another class of van der Waals material known as the transition metal dichalcogenides has also been predicted to be compelling for hybrid quantum systems\cite{gong2013magnetoelectric, schaibley2016valleytronics, bussolotti2018roadmap, goh2020toward, goh2023valleytronics}. Because of the possibility of being direct bandgap semiconducting materials in the monolayer form as well as possessing valley protected spin states for robust qubits, these materials are naturally optoelectronics compatible on the same material platform, and the relative ease of stacking these materials to form heterostructures opens even more degress of freedom for not just photonic, but also excitonic and trionic devices that offer rich possibilities for different quantum modalities to coexist and interact on the same platform. In addition, the inherent valleypseudospin in such materials offers the valley index as a good quantum number for state information, thus making a pure valley qubit\cite{kormanyos2014spin} an alluring prospect which might benefit from hybrid phononic coupling for manipulation in the momentum space. Although these are very new materials, recent significant progress in their processing and integration into devices augurs well for them to be next generation contenders for hybrid quantum devices.

\vspace{5mm} 

The field of hybrid quantum systems is nascent even for the more focused sub-field of artificial atoms. Nevertheless, the beginning of systematic categorization rationalized upon principles of new functionalities, practicality and integrability should be a continuous endeavor to guide research and development. This should hopefully lead to more rapid translation into useful applications and guidance for standardization of quantum technologies for interoperability.

\vspace{5mm} 

\medskip
\textbf{Acknowledgements} \par 
C.C. acknowledges funding support from Agency for Science, Technology and Research MTC YIRG Grant M22K3c0107. D.H. acknowledges funding support from Agency for Science, Technology and Research MTC YIRG Grant M22K3c0105 and Agency for Science, Technology and Research Grant C222812022. V.L. acknowledges funding support from Agency for Science, Technology and Research Grant C230917005. J.F.K. acknowledges funding support from Agency for Science, Technology and Research Grant C230917003. K.E.J.G. acknowledges the funding support from the Agency for Science, Technology and Research Grant C230917006 and the Singapore National Research Foundation Grants CRP21-2018-0001 and NRF2021-QEP2-02-P07.

\medskip

%
\bibliographystyle{MSP}
\bibliography{main_ref}

\end{document}